\documentclass[3p,times,procedia]{elsarticle}
\usepackage{nupha_ecrc}

\volume{00}

\firstpage{1}

\journalname{Nuclear Physics A}

\runauth{}

\jid{nupha}

\jnltitlelogo{Nuclear Physics A}

\usepackage{amssymb}

\usepackage[figuresright]{rotating}

\begin{document}

\begin{frontmatter}



\dochead{XXVIIIth International Conference on Ultrarelativistic Nucleus-Nucleus Collisions\\ (Quark Matter 2019)}







\title{Rotation Effects on Mesonic Condensation in Isospin Matter}
\author[1,2,3] {Hui Zhang}
\author[1] {Defu  Hou}
\ead{houdf@mail.ccnu.edu.cn , \,\,  Presenter}
\author[3,1]{ Jinfeng Liao}
\address[1]{ Institute of Particle Physics (IOPP) and Key Laboratory of Quark and Lepton Physics (MOE),  Central China Normal University, Wuhan 430079, China}
\address[2]{Guangdong Provincial Key Laboratory of Nuclear Science, Institute of Quantum Matter, South China Normal University, Guangzhou 510006, China}
\address[3]{ Physics Department and Center for Exploration of Energy and Matter, Indiana University, 2401 N Milo B. Sampson Lane, Bloomington, IN 47408, USA.}

\date{\today}

\begin{abstract}
We investigate the rotation effects on the  mesonic condensation in isospin matter . Using the two-flavor NJL effective model with a  global rotation, we  demonstrate two important effects of the rotation on its phase structure: a rotational suppression of the scalar-channel condensates, in particular the pion condensation region; and a rotational enhancement of the rho condensation region with vector-channel condensate. A new phase diagram for isospin matter under rotation is   mapped out on the $\omega-\mu_I$ plane  where three distinctive phases, corresponding to $\sigma$, $\pi$, $\rho$ dominated regions respectively, are separated by a second-order line at low isospin chemical potential and a first-order line at high rotation which are further connected at  a tri-critical point.

  \end{abstract}

\end{frontmatter}




\newcommand{\beq}{\begin{equation}}
\newcommand{\eeq}{\end{equation}}
\newcommand{\ret}{\mathrm{ret}}
\newcommand{\adv}{\mathrm{adv}}



\section{\label{sec:Intro} Introduction}

Recently there have been rapidly growing interests in understanding the properties and phase structures of matter under extreme fields like magnetic field or global rotation~\cite{Miransky:2015ava,Fukushima:2018grm}.
Rotation provides an interesting new type of macroscopic control parameter, in addition to conventional ones such as temperature and density, for a many-body system. In particular it has nontrivial interplay with microscopic spin degrees of freedom through the rotational polarization effect and could often induce novel phenomena. For example, there are highly nontrivial anomalous transport effects such as the chiral vortical effect and chiral vortical wave in rotating fluid with chiral fermions~\cite{Son:2009tf,Kharzeev:2010gr,Landsteiner:2011iq,Hou:2012xg,Jiang:2015cva,Flachi:2017vlp,Kharzeev:2015znc}. Furthermore, if the underlying materials contain fermions that may form condensate via pairing, their phase structure can be significantly influenced by the presence of global rotation~\cite{Jiang:2016wvv,Ebihara:2016fwa,Chen:2015hfc,Mameda:2015ria,Huang:2017pqe,Liu:2017spl}. A generic effect is the rotational suppression of  fermion pairing in the zero angular momentum states, which has been demonstrated for e.g. chiral phase transition and color superconductivity in the strong interaction  system~\cite{Jiang:2016wvv}.

In this report, we perform the first analysis on the influence of rotation on the phase structure of isospin matter --- the Quantum Chromodynamics (QCD) matter at finite isospin density (or equivalently chemical potential)\cite{Son:2000xc,Son:2000by}.
We will perform the first systematic study of all possible mesonic condensation pairing states simultaneously in the isospin matter under global rotation.  We will show that there are  suppression of scalar pairing and enhancement of vector pairing due to fluid rotation, with the emergence of rho condensation phase at high isospin density under rapid rotation. Such analysis will further allow us to envision and map out a new phase diagram on the  rotation-isospin   parameter plane with highly nontrivial phase structures.

To investigate the mesonic condensation in isospin matter, we will adopt a widely-used effective model, namely the two-flavor  Nambu-Jona-Lasinio (NJL) model with four-fermion interactions in various channels at finite isospin chemical potential $\mu_I$:
\begin{equation} \label{eq_1}
{\cal L} = \bar\psi(i\gamma_\mu\partial^\mu-m_0+{\mu_I\over2}\gamma_0\tau_3)\psi
+  {G_s}\left[\left(\bar\psi\psi\right)^2+\left(\bar\psi i\gamma_5{\bold \tau}\psi\right)^2\right]  -{G_v} \left(\bar\psi \gamma_\mu{\bold \tau}\psi\right)^2  \,\, .
\end{equation}
In the above, the $m_0= 5\rm MeV$ is the light quark mass parameter while $G_s=G_v=5.03\ {\rm GeV^{-2}}$ are the scalar and vector channel coupling constants respectively. The NJL-type effective model also requires a momentum cut-off parameter  $\Lambda=650\ {\rm MeV}$. These choices are quite standard, leading to the correct pion mass and decay constant in the vacuum as well as a vacuum expectation value (VEV) of $\sigma$ field to be $\sigma_0=2\times(250\rm MeV)^3$.

In the most general case, we consider three possible mesonic condensation scenarios: condensation of $\sigma$, $\pi$ or $\rho$ fields respectively.  Following the standard mean-field method, we introduce the corresponding condensates:
$ \sigma= \langle \bar\psi\psi \rangle,\
\pi= \langle \bar\psi i \gamma_5 {\tau_3}  \psi \rangle,\
\rho= \langle \bar\psi i \gamma_0 {\tau_3} \psi \rangle$.

Furthermore we are considering such a system under global rotation around $\hat{z}$-axis with angular velocity $\vec{\omega}=\omega \hat{z}$. To do this, one can study the system in the rotating frame and rewrite the spinor theory with the curved metric associated with the rotating frame~\cite{Jiang:2016wvv}.  In such a description, the main new effect is a global polarization term in the Lagrangian density:
${\cal L}_{R} = \psi^\dagger \left[  (\vec{\omega}\times \vec{x})\cdot (-i \vec{\partial})
+ \vec{\omega} \cdot \vec{S}_{4\times 4} \right] \psi$
where $\vec{S}_{4\times 4} = \frac{1}{2} Diag\left(\vec{\sigma} , \vec{\sigma} \right)$ is the spin operator with $\vec{\sigma}$ the $2\times 2$ Pauli matrices.
Physically, this term polarizes  both the orbital and spin angular momenta to be aligned with global rotation axis and its effect is identical for particles or antiparticles.

Within mean-field approximation and assuming $\omega\ r \ll 1$, one can obtain  thermodynamic potential of NJL model for isospin matter under rotation $\Omega$ \cite{Hui2019}.
At given temperature $T$ and isospin chemical potential $\mu_I$, one then determines the mean-field condensates by  solving the gap equations:
\begin{eqnarray} \label{eq_Gap}
\frac{\partial \Omega}{\partial \sigma}=\frac{\partial \Omega}{\partial \pi}=\frac{\partial \Omega}{\partial \rho}=0 ,
\end{eqnarray}
These equations can be numerically solved. For situations with multiple solutions,  the true physical state should be determined from the absolute minimum of the thermodynamic potential. For the numerical results to be presented later, we use a value $r=0.1\ {\rm GeV}^{-1}$.

\section{\label{sec:Scalar} Rotational Suppression of Pion Condensation}

We first demonstrate the rotational suppression of scalar pairing channels, i.e. mesonic condensates arising from quark-anti-quark pairing in the zero total angular momentum $J=0$ states.  While such rotational suppression was previously proposed as a generic phenomenon in fermion pairing transitions and demonstrated for e.g. chiral condensate or color superconductivity~\cite{Jiang:2016wvv}, it has never been examined for the mesonic condensation in isospin matter. In our case, the scalar pairing channels include the condensates of both $\sigma$ (scalar) and $\pi$ (pseudo-scalar) fields. To show this effect clearly, we will temporarily ``turn off'' the vector channel in the present section.
\begin{figure}[htb!]
\centering
\includegraphics[width=140pt]{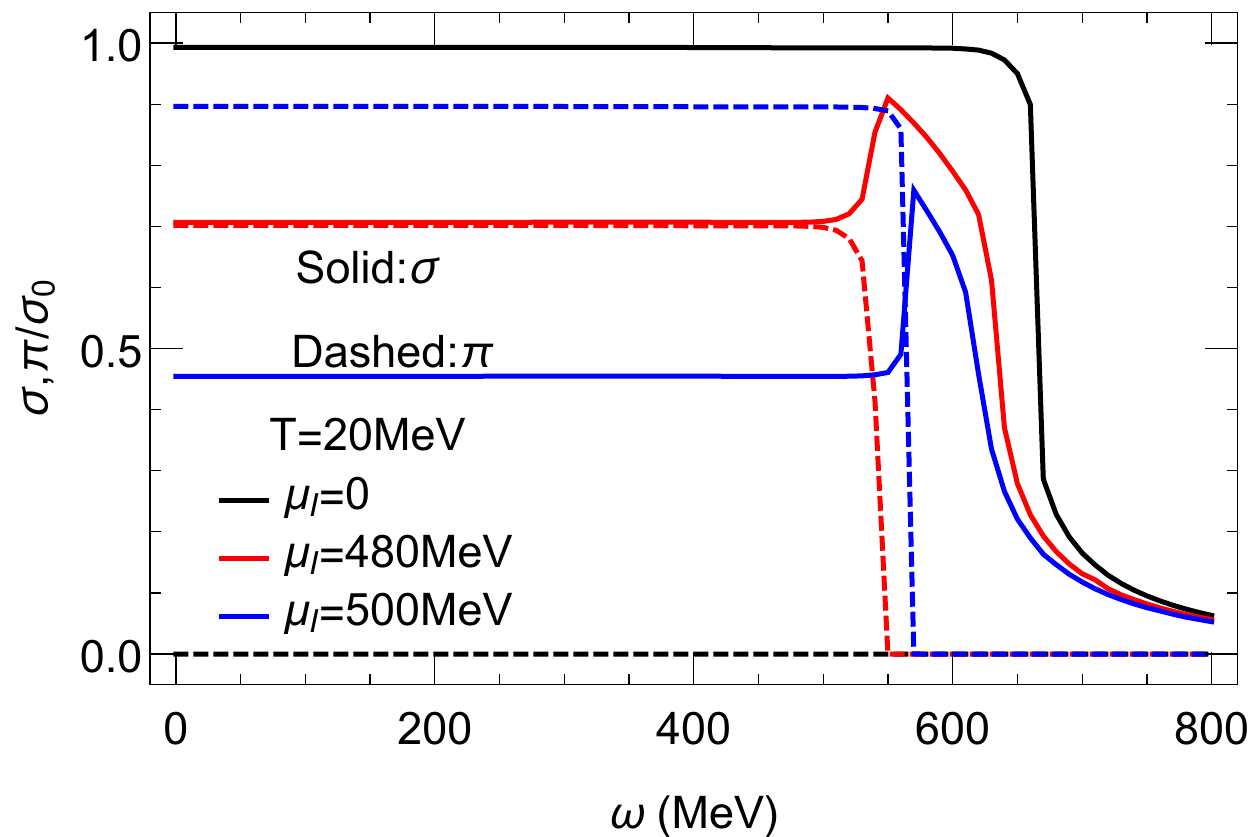}
\includegraphics[width=140pt]{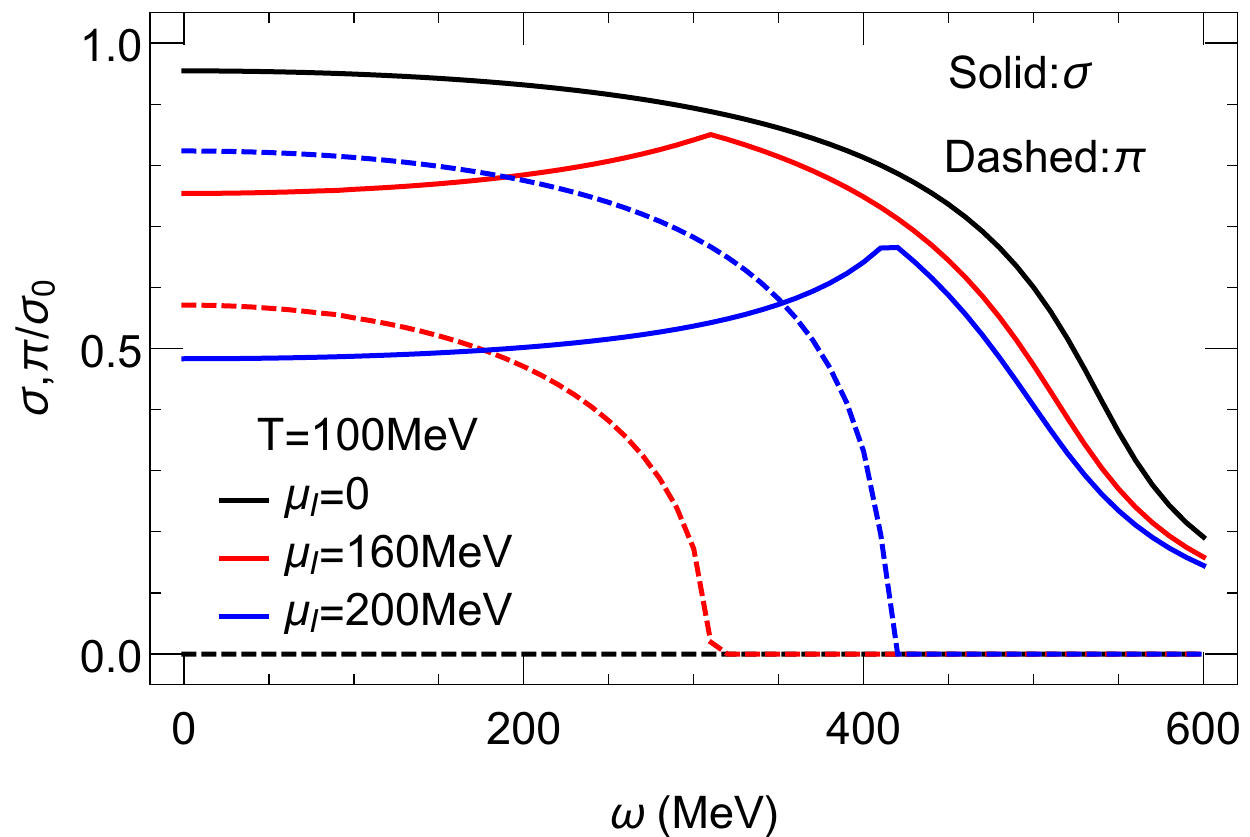}
\label{fig1}
\caption{ (color online)  $\sigma$ and $\pi$ condensates as a function of $\omega$ at $T=20\rm MeV$  and $T=100\rm MeV$ (right) for  different values of $\mu_I$.}
\end{figure}
In Fig.~\ref{fig1}, we show the sigma and pion condensates $\sigma$ and $\pi$ (scaled by $\sigma_0$) as a function of $\omega$ at $T=20\rm MeV$  and $T=100\rm MeV$ for several different values of $\mu_I$. As one can see, there is indeed a generic suppression on both the $\sigma$ and $\pi$ condensates, due the fact that the presence of global rotation always ``prefers'' states with nonzero angular momentum and thus disfavors these $J=0$ mesonic pairing channels. What's most interesting is the case at high isospin density, where the system is in a pion condensation phase with nonzero pion condensate without rotation. But with increasing rotation, this condensate eventually approaches zero via either a first-order (at low T) or second-order (at high T) transition. Thus the spontaneously broken isospin symmetry in the pion condensation phase can be restored again under rapid rotation, which is a new effect.

\section{\label{sec:Vector} Enhanced Rho Condensation under Rotation}

\begin{figure}[htb!]
\includegraphics[width=130pt]{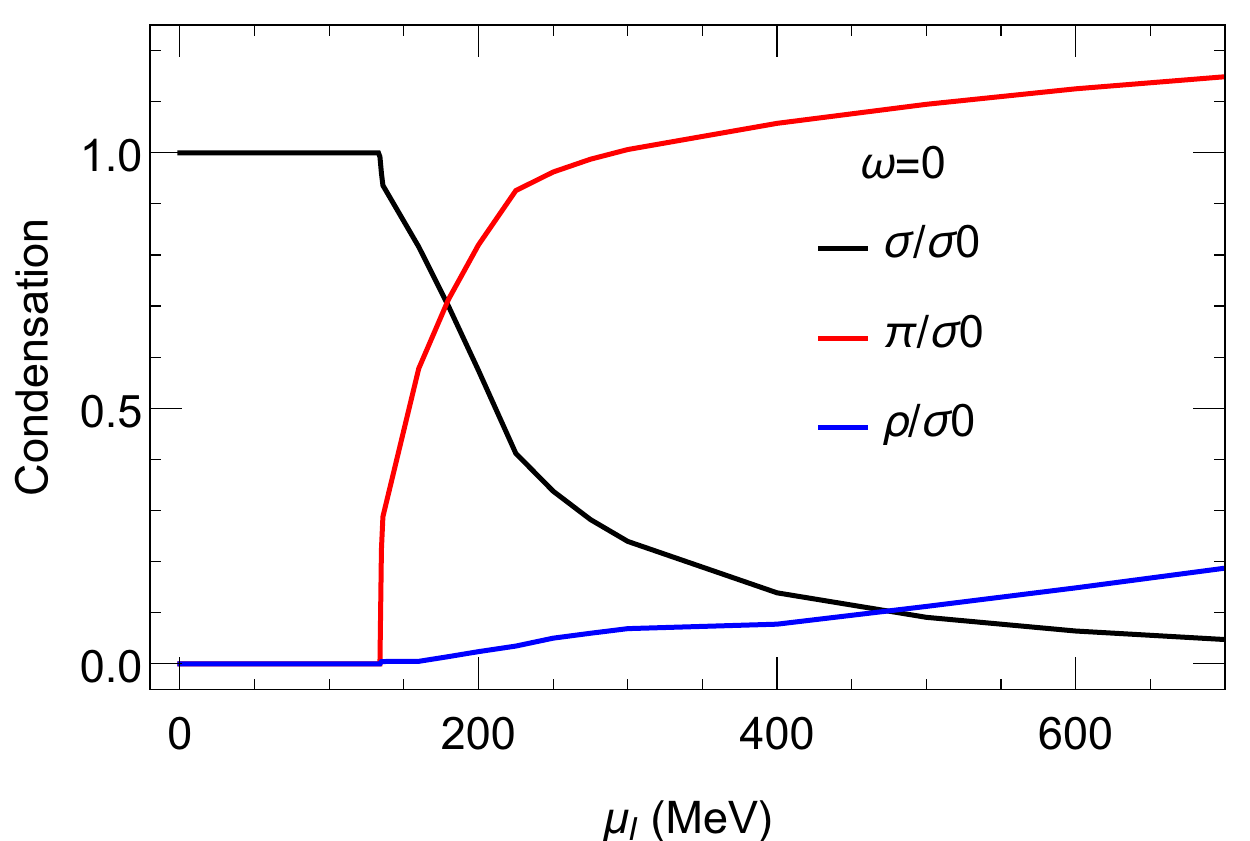}
\includegraphics[width=130pt]{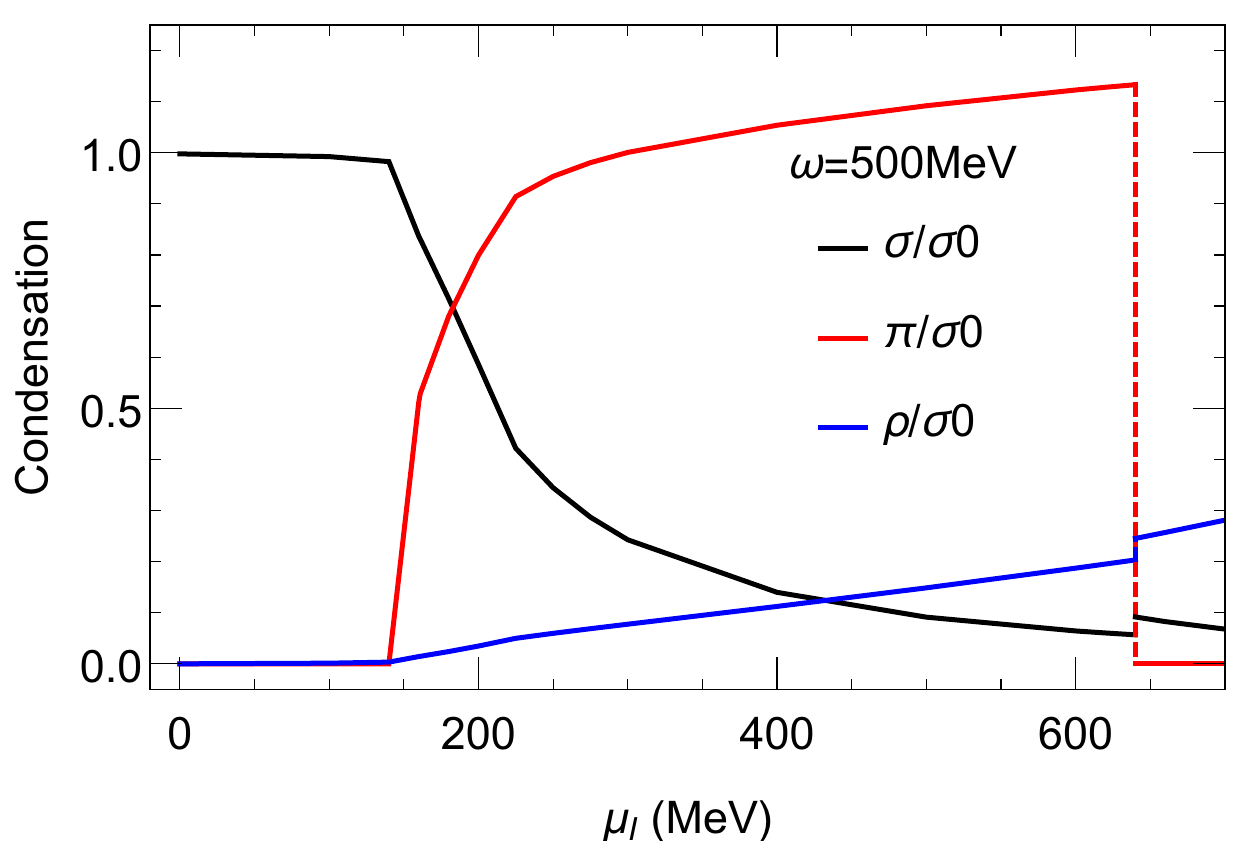}
\includegraphics[width=130pt]{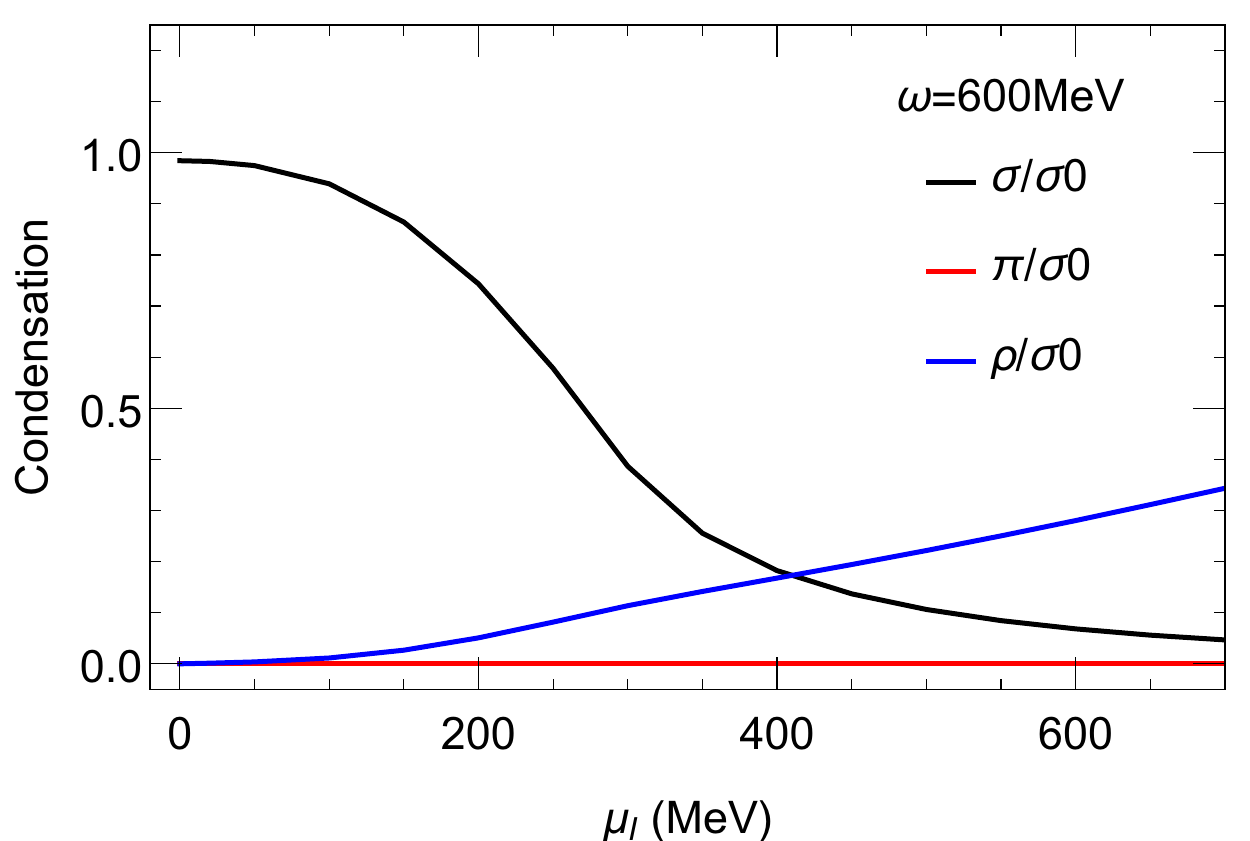}
\caption{ (color online) The sigma, pi and rho condensates $\sigma,\ \pi,\ \rho$ (scaled by the vacuum chiral condensate $\sigma_0$) as a function of isospin chemical potential, for
$\omega=0$ , $\omega=500\rm MeV$ (middle) and $\omega=600\rm MeV$, respectively. The temperature is $T=10\rm MeV$.}\label{fig3}
\end{figure}

Suppression of the scalar pairing implies opportunity for enhanced pairing of states with nonzero angular momentum, such as the $\rho$ condensate. Indeed, the $\rho$ state has $J=1$ and should be favored by the presence of global rotation. While the emergence of $\rho$ condensate at high isospin density has been previously studied~\cite{Brauner:2016lkh}, the interplay between the rho condensate and rotation and the implication for phase structure of isospin matter is discussed for the first time here.
 To do this, we now consider the full thermodynamic potential and consistently solve the coupled gap equations of all three possible condensates . In Fig.\ref{fig3}, we compare the results for the sigma, pi and rho condensates $\sigma,\ \pi,\ \rho$ (scaled by the vacuum chiral condensate $\sigma_0$) as a function of isospin chemical potential, for $\omega=0$ (upper), $\omega=500\rm MeV$ (middle) and $\omega=600\rm MeV$, respectively. The temperature for this calculation is $T=10\rm MeV$.

In the case without rotation (Fig.\ref{fig3} left panel), the chiral condensate decreases with increasing $\mu_I$  while both pion
and rho condensates start to grow for $\mu_I$ greater than the critical value at about $140\rm MeV$ for a second order phase
transition.  The pion condensate dominates the system at large isospin chemical potential.

In the case with strong rotation, $\omega=500\rm MeV$ (Fig.\ref{fig3} middle panel),  the situation becomes different. Both pion and rho condensates still start to grow for $\mu_I$ greater than the critical value. But at even higher isospin density, a new first-order transition occurs and the pion condensate drops to zero. In this new region, the rho condensate becomes dominant.

For even stronger rotation,  $\omega=600 \rm MeV$ (Fig.\ref{fig3} right panel), the pion condensate disappears all together. With increasing isospin chemical potential $\mu_I$, there is a smooth crossover from a $\sigma$-dominated phase at low isospin density to a
$\rho$-dominated phase at very high isospin density.

\begin{figure}[htb!]
\centering
\includegraphics[width=140pt]{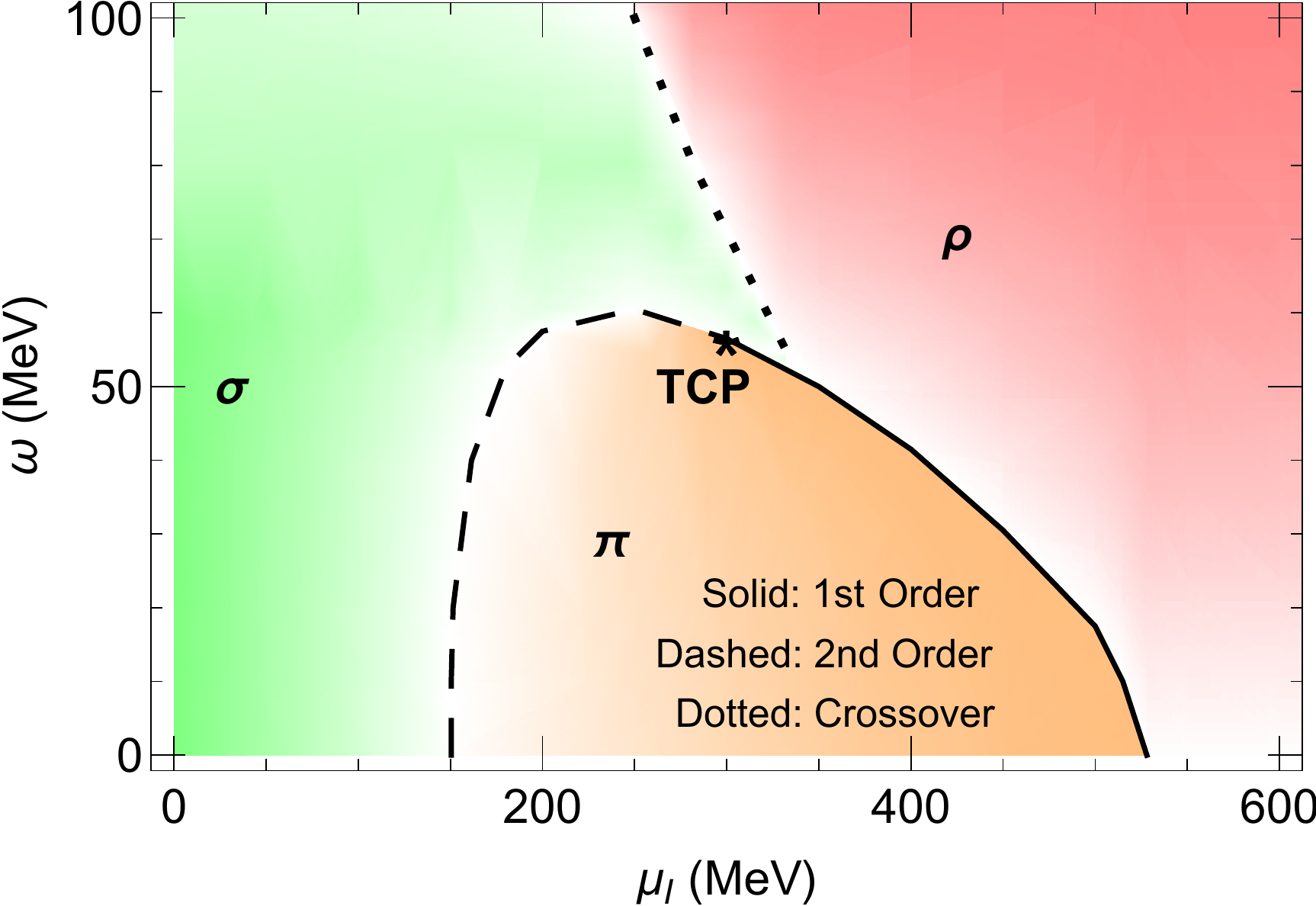}
\caption{ (color online) A new phase diagram on the $\omega-\mu_I$ plane for mesonic condensation in isospin matter under rotation.   solid line stands for 1st-order phase transition and dashed line  for 2nd-order transition, while dotted line for crossover, with the star symbol denoting a tri-critical point (TCP) at  $(\mu_I^c=165{\rm MeV},\omega^c=548{\rm MeV})$. The temperature is $T=10\rm MeV$ and  $\mu_B=250\rm MeV$.}\label{fig4} \vspace{-0.2in}
\end{figure}

These results clearly demonstrate the influence of rotation on the mesonic condensation in isospin matter and envision a new phase diagram on the $\omega-\mu_I$ plane, as shown in Fig.~\ref{fig4}. This new phase structure is characterized by three distinctive regions: a vacuum-like, sigma-dominated phase in the low isospin density and slow rotation region; a pion-condensation phase in the mid-to-high isospin density with moderate rotation; and a rho-condensation phase in the high isospin and rapid rotation region. A second-order transition line   separates the sigma-dominant and pion-dominant regions while a first-order  line   separates the pion-dominant and rho-dominant regions, with a tri-critical point connecting them. Similar phase structure has been found  at various temperatures and baryon chemical potentials, and is a robust feature from interplay between rotation and isospin.

\section{\label{sec:summary} Conclusion}

In this paper, we have investigated the mesonic condensation in isospin matter under rotation. Using the two-flavor NJL effective model under the presence of global rotation, we have demonstrated two important effects of the rotation on the phase structure: a rotational suppression of the scalar-channel condensates, in particular the pion condensation region; and a rotational enhancement of the rho condensation region with vector-channel condensate. A new phase diagram for isospin matter under rotation has been mapped out on the $\omega-\mu_I$ plane where three distinctive phases, corresponding to $\sigma$, $\pi$, $\rho$ dominated regions respectively, are separated by a second-order line at low isospin chemical potential and a first-order line at high rotation which are joint by a tri-critical point. While the quantitative details    may depend on model details, we expect such a three-region phase structure to be generic.

  {\em Acknowledgements.---}
  This work is in part supported by NSFC Grant Nos. 11735007,11890711,  by NSF Grant No. PHY-1352368, PHY-1913729 and by the U.S. Department of Energy, Office of Science, Office of Nuclear Physics, within the framework of the Beam Energy Scan Theory (BEST) Topical Collaboration. HZ acknowledges partial support from the China Scholarship Council. JL is  grateful to the Institute for Advanced Study of Indiana University for partial support.

\vspace{-0.2in}


\end{document}